\documentclass[superscriptaddress,twocolumn]{revtex4}
\usepackage[tbtags]{amsmath}
\usepackage{amsfonts}
\usepackage{mathrsfs}
\usepackage{amssymb}
\usepackage{graphicx}
\usepackage{epsfig,graphicx,times}
\usepackage{subfigure}
\usepackage{color}
\usepackage{CJK}

\begin{document}

\title{Quantum routing of single photons with whispering-gallery resonators}

\author{Jin-Song Huang\footnote{jshuangjs@126.com} }

\affiliation{School of Information Engineering, Jiangxi University
of Science and Technology, Ganzhou 341000, China}

\author{Jia-Hao Zhang}

\affiliation{School of Information Engineering, Jiangxi University
of Science and Technology, Ganzhou 341000, China}

\author{L. F. Wei\footnote{weilianfu@gmail.com}}
\affiliation{Information Quantum Technology Laboratory, School of Information Science
and Technology, Southwest Jiaotong University, Chengdu 610031,
China}
\affiliation{State Key Laboratory of Optoelectronic Materials
and Technologies, School of Physics, Sun Yat-Sen
University Guangzhou 510275, China}

\begin{abstract}
Quantum routing of single photons in a system with two waveguides
coupled to two whispering-gallery resonators (WGRs) are investigated
theoretically. With a real-space full quantum theory, photonic
scattering amplitudes along four ports of the waveguide network are
analytically obtained. It is shown that, by adjusting the geometric
and physical parameters of the two-WGR configuration, the quantum
routing properties of single photons along the present waveguide
network can be controlled effectively. For example, the routing
capability from input waveguide to another one can significantly
exceed 0.5 near the resonance point of scattering spectra, which can
be achieved with only one resonator. By properly designing the
distance between two WGRs and the waveguide-WGR coupling strengths,
the transfer rate between the waveguides can also reach certain
sufficiently high values even in the non-resonance regime. Moreover,
Fano-like resonances in the scattering spectra are designable. The
proposed system may provide a potential application in controlling
single-photon quantum routing as a novel router.


\end{abstract}

\maketitle

\section{Introduction}

In the realization of various quantum network~\cite{Kimble}, quantum
nodes, coherently connecting different quantum channels, are
particularly important ingredients. Quantum routers~\cite{Aoki,Hoi},
as one kind of node devices, are usually utilized to control the
path of the quantum signals. Recent development of the single-photon
transport technique~\cite{Shena,Shenb,Chang,Shen,Zhoua} provides an
ideal tool to design various desired quantum networks, wherein
photonic waveguides are served as the quantum channels for
propagating singles and the quantum routers are utilized to control
the single photons in the channels. Indeed, many
theoretical~\cite{Zhou,Lu,Yan,Lu2,Agarwal,Xia,Lemr} and
experimental~\cite{Ma,Shomroni} works have demonstrated
how to design various quantum routers for controlling the photonic
transports in quantum networks.

Typically, Zhou and Lu et al~\cite{Zhou,Lu} proposed a routing
approach to control the photonic propagations in the X-shaped
coupled-resonator waveguide. Also, single-photon routing scheme for
the usual optical waveguide with multiple ports has been proposed by
Yan et al~\cite{Yan}. However, the expectable routing probabilities
in these configurations are limited no more than 0.5 by a single
two-level atom or a single bosonic mode, and thus restrict more
practical applications. To overcome partially such a difficulty, we
proposed an alternative scheme~\cite{Wei} to implement various
designable routing probabilities for arbitrarily selected ports, by
using a series of atomic mirrors. Still, the routing capability of
the single photon controlled by the atomic mirrors is strongly
influenced by their unavoidable dissipations. Therefore, looking for
the robust quantum router to implement the single-photon routing
with high probability in multiple waveguide network is still a
challenge and thus of considerable interest.

It is well-known that the whispering-gallery resonator (WGR), which
supports two bosonic modes, is one of the attractive optical
devices. Particularly, due to its very low loss rate achieved
experimentally and the special construction of its cyclic modes, the
WGR possesses the potential advantage for many quantum optical
applications. Actually, with the WGRs, numerous single-photon
devices such as quantum switches~\cite{Nozaki,Poon,Shen0,Schmid},
photonic transistors~\cite{Hong,Hong1}, quantum routing
networks~\cite{Kimble,Yao}, entangled photon-pair
generators~\cite{Ajiki}, and optomechanical devices~\cite{Safavi,
Groeblacher,Connell,Weis,Rocheleau}, etc., have been demonstrated.
Continuously, in this paper we design a double-WRG quantum router to
control the single-photon transport in a double-waveguide quantum
network.

By adopting a full quantum theory on single-photon transports in
real space~\cite{Shen}, we investigate how the single-photon can be
controlled in the waveguide network by using the scatterings by a
pair of WRGs with the inter-resonator distance and the
resonator-waveguide coupling strength being designable. The
amplitudes of single-photon being scattered into the four ports of
the network are obtained analytically. Then, by numerical method we
discuss how to design the relevant geometric and physical parameters
to implement the desired quantum routings of the single photons. It
is shown that the quantum routing property of the single photons
into these four ports can be controlled really by adjusting
backscattering strength between the two degenerate mode of WGR. In
contrast to previous schemes~\cite{Zhou,Lu,Yan} with a maximum
transfer rate $0.5$, the probability of single photons routed from
input channel to another can significantly exceed 0.5 near resonance
even for a single resonator, due to the special cyclic mode of the
resonator to redirect the photons to another waveguide. High routing
capability can also be realized by varying the distance of two WGRs
and the coupling between waveguides and WGRs in the nonresonance
regime. Additionally, we find that the Fano-like resonances, due to
the quantum interference, are also be designed for its potential
applications.

The paper is organized as follows. Sec. II shows our theoretical
model and its exact solution. Accordingly in Sec.~III, by numerical
method we investigate how to adjust double-WGR router parameters to
implement high routing capabilities of single photons, and discuss
the controls of the Fano-like resonances in the present
configuration. Our conclusions are summarized finally in Sec.~IV.

\section{Theoretical mode}

The considered waveguide network is illustrated schematically in
Fig.~1, wherein two WGRs coupled to two waveguides are separated
with a distance $d$. The waveguides are utilized as the quantum
channels, and each of the WGRs supports two degenerate modes, i.e.,
the clockwise- and counterclockwise ones. The black lines and red
lines denote the counterclockwise and clockwise propagations of the
incident photon, respectively. $R_{m}$ and $T_{m}(m=a,b)$ represent
the reflection and transmission amplitudes in the $m$th-waveguide of
the single photons, which is incident from the left of the
waveguide-a and routed by the WGRs into the four ports of the
network.

\begin{figure}[htbp]
\includegraphics[width=8cm,height=5cm]{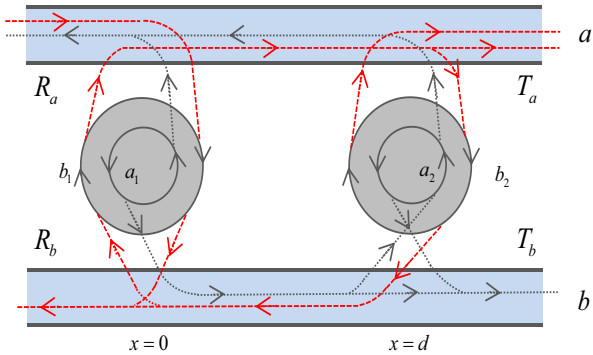}
\caption{(Color online) Schematic routings of single photons in two
photonic waveguides (labeled as a and b, respectively), controlled
by two WGRs with respective degenerate modes, $a_{n}$ and
$b_{n}(n=1,2)$. Here, the red and the black lines represent the
clockwise and counterclockwise propagations, respectively. $R_{a}$,
$R_{b}$ and $T_{a}$, $T_{b}$ are the reflection and transmission
amplitudes of the photons, respectively.}
\end{figure}

Suppose that the photonic dispersions in the waveguide~\cite{Shen}
are linear. Then the effective real-space Hamiltonian for the
present network can be written as
\begin{eqnarray}
H/\hbar&=&\int dx {\bigg{\{}}
\sum_{m=a,b}[-i\upsilon_{g}\frac{\partial}{\partial
x}C_{Rm}^{\dag}(x)C_{Rm}(x)\nonumber\\
&+&i\upsilon_{g}\frac{\partial}{\partial
x}C_{Lm}^{\dag}(x)C_{Lm}(x)] \nonumber\\
&+&V_{1a}\delta(x)[C_{Ra}^{\dag}(x)b_{1}+C_{La}^{\dag}(x)a_{1}+H.c.]\nonumber\\
&+&V_{1b}\delta(x)[C_{Rb}^{\dag}(x)a_{1}+C_{Lb}^{\dag}(x)b_{1}+H.c.]\nonumber\\
&+&V_{2a}\delta(x-d)[C_{Ra}^{\dag}(x)b_{2}+C_{La}^{\dag}(x)a_{2}+H.c.]\nonumber\\
&+&V_{2b}\delta(x-d)[C_{Rb}^{\dag}(x)a_{2}+C_{Lb}^{\dag}(x)b_{2}+H.c.]{\bigg{\}}}\nonumber\\
&+&\sum_{n=1,2}^{m=a,b}[(\omega_{cn}-\frac{i\gamma_{cn}}{2})m_{n}^{\dag}m_{n}+h_{n}(a_{n}^{\dag}b_{n}
+b_{n}^{\dag}a_{n})].\nonumber\\
\label{eq:ham}
\end{eqnarray}
Here, $C_{Rm}^{\dag}(x)$/$C_{Lm}^{\dag}(x)$ denotes the creation
operator of the right/left-moving photon at $x$ in the waveguide-m.
$\upsilon_g$ is the group velocity of the photons and $V_{nm}(n=1,
2; m=a, b$ throughout the paper) the coupling strength between the
$n$th WGR and the waveguide-m. $\delta(x)/\delta(x-d)$ indicates
that the interaction between the waveguide and the WGR occurs at
$x=0/d$. Also, $a_{n}^\dagger(a_{n})$ is the bosonic
creation(annihilation) operator for the counterclockwise mode of the
$n$th WGR, while $b_{n}^\dagger(b_{n})$ for the clockwise mode of
the $n$th WGR with the eigenfrequency $\omega_{cn}$. $\gamma_{cn}$
is the dissipation rates of the $n$th WGR mode, and $h_{n}$ is the
backscattering strength between the clockwise mode and
counterclockwise mode, respectively.

Assume that the photon is incident from the left in the waveguide-a
with the energy $E_k =\upsilon_{g}k$. The scattering eigenstates of
the Hamiltonian (\ref{eq:ham}) are given by
\begin{eqnarray}
|\psi\rangle &=&\sum_{m=a,b} \int  dx[ \phi_{Rm}(x) C_{Rm}^\dag(x) \nonumber\\
&+& \phi_{Lm}(x)
C_{Lm}^\dag(x)]|\emptyset_{a},\emptyset_{b},0,0,0,0\rangle\nonumber\\
&+&\xi_{1}|\emptyset_{a},\emptyset_{b},1,0,0,0\rangle+\xi_{2}|\emptyset_{a},\emptyset_{b},0,1,0,0\rangle\nonumber\\
&+&\xi_{3}|\emptyset_{a},\emptyset_{b},0,0,1,0\rangle+\xi_{4}|\emptyset_{a},\emptyset_{b},0,0,0,1\rangle
\end{eqnarray}
where $\phi_{Rm/Lm}$ is the single photon wave function in the
right/left of the waveguide-m. $|\emptyset_{a},\emptyset_{b}, 0, 0,
0, 0\rangle$ describes the vacuum states of the waveguides and the
WGRs. $\xi_1$ and $\xi_{2}$ are the excitation amplitude of
clockwise and counterclockwise modes of the WGR-1. $\xi_{3}$ and
$\xi_{4}$ are the excitation amplitudes of clockwise and
counterclockwise modes of the WGR-2. Furthermore, the above
amplitudes can be expressed formally as
\begin{eqnarray}
&&\phi_{Ra}(x)=e^{ikx}[\theta(-x)+t_{12}^{a}\theta(x)\theta(d-x)+t_{a}\theta(x-d)],\nonumber\\
&&\phi_{La}
(x)=e^{-ikx}[r_{a}\theta(-x)+r_{12}^{a}\theta(x)\theta(d-x)],\nonumber\\
&&\phi_{Rb}(x)=e^{ikx}[t_{12}^{b}\theta(x)\theta(d-x)+t_{b}\theta(x-d)],\nonumber\\
&&\phi_{Lb}(x)=e^{-ikx}[r_{b}\theta(-x)+r_{12}^{b}\theta(x)\theta(d-x)].
\end{eqnarray}
where $t_a$ and $r_a$ are the transmission and reflection amplitudes
in the waveguide-a, $t_b$ and $r_b$ are the transmission and
reflection amplitudes in the waveguide-b, respectively. $\theta(x)$
is the Heaviside step function with $\theta(0)=1/2$.
$t_{12}^{a(b)}\theta(x)\theta(d-x)$ and
$r_{12}^{a(b)}\theta(x)\theta(d-x)$ represent the transferred
amplitudes between the WGRs.

Solving the eigen-equation $H|\psi\rangle=E_k|\psi\rangle$, the expressions of $t_a$, $r_a$, $t_b$ and $r_b$ can be
obtained as follows:
\begin{eqnarray}
t_{a}&=&1+\frac{\Gamma_{1a}Q_{2}+\Gamma_{2a}Q_{1}-iB_{a}[u_{1}e^{i\phi(k)}+u_{2}]}{i[Q_{1}Q_{2}+u_{1}u_{2}e^{i\phi(k)}]},\nonumber\\
r_{a}&=&-\frac{iB_{b}S_{1}e^{i\phi(k)}[B_{a}Q_{2}-iu_{1}\Gamma_{2a}e^{i\phi(k)}]
}{i[Q_{1}Q_{2}+u_{1}u_{2}e^{i\phi(k)}]}\nonumber\\ &+&\frac{A_{1}S_{2}e^{i\phi(k)}(\Gamma_{2a}Q_{1}-iB_{a}u_{2})}{i[Q_{1}Q_{2}+u_{1}u_{2}e^{i\phi(k)}]}\nonumber\\
&-&\frac{iB_{a}S_{2}e^{i\phi(k)}(B_{a}Q_{1}-iu_{2}\Gamma_{1a})
}{i[Q_{1}Q_{2}+u_{1}u_{2}e^{i\phi(k)}]}
\nonumber\\ &+&\frac{A_{2}S_{1}[\Gamma_{1a}Q_{2}-iB_{a}u_{1}e^{i\phi(k)}]}{i[Q_{1}Q_{2}+u_{1}u_{2}e^{i\phi(k)}]},\nonumber\\
t_{b}&=&-\frac{iB_{b}S_{1}[Q_{2}D_{1}-iu_{1}C_{2}e^{i\phi(k)}]}{i[Q_{1}Q_{2}+u_{1}u_{2}e^{i\phi(k)}]}
\nonumber\\ &+&\frac{A_{1}S_{2}(Q_{1}C_{2}-iu_{2}D_{1})}{i[Q_{1}Q_{2}+u_{1}u_{2}e^{i\phi(k)}]}\nonumber\\
&-&\frac{iB_{a}S_{2}e^{i\phi(k)}(Q_{1}D_{2}-iu_{2}C_{1})}{i[Q_{1}Q_{2}+u_{1}u_{2}e^{i\phi(k)}]}
\nonumber\\ &+&\frac{A_{2}S_{1}[Q_{2}C_{1}-iu_{1}D_{2}e^{i\phi(k)}]}{i[Q_{1}Q_{2}+u_{1}u_{2}e^{i\phi(k)}]},\nonumber\\
r_{b}&=&\frac{Q_{1}C_{2}e^{i\phi(k)}+Q_{2}C_{1}-ie^{i\phi(k)}(u_{2}D_{1}+u_{1}D_{2})}{i[Q_{1}Q_{2}+u_{1}u_{2}e^{i\phi(k)}]},
\nonumber\\
 \label{eq:tr}
\end{eqnarray}
where $E_{kn}=E_{k}-(\omega_{cn}-i\gamma_{cn}/2)$,
$\Gamma_{nm}={V_{nm}^{2}}/{\upsilon_{g}}$,
$A_{n}=E_{kn}+i(\Gamma_{na}+\Gamma_{nb})/2$,
$B_{m}=\sqrt{\Gamma_{1m}\Gamma_{2m}}$,
$C_{n}=\sqrt{\Gamma_{na}\Gamma_{nb}}$,
$D_{1}=\sqrt{\Gamma_{1a}\Gamma_{2b}}$,
$D_{2}=\sqrt{\Gamma_{2a}\Gamma_{1b}}$, $\phi(k)=2kd$,
$M=A_{1}A_{2}+B_{1}B_{2}e^{\phi(k)}$, $S_{n}=h_{n}/M$,
$Q_{1}=A_{1}-A_{2}h_{1}S_{1}$, $Q_{2}=A_{2}-A_{1}h_{2}S_{2}$,
$u_{1}=B_{2}+B_{1}h_{1}S_{2}$, and $u_{2}=B_{1}+B_{2}h_{2}S_{1}$.

Analytic expressions demonstrated above provide a complete
description on the single-photon transport properties of the
proposed network. Obviously, by properly designing the relevant
geometric parameter $d$ and the other physical parameters, such as
the frequency of the incident photons, the waveguide-resonator
coupling strengths $V_{nm}$, and the intermode interactions $h_n$,
the desired photon routings can be implemented.

\section{Quantum routings of single photons by engineering the WGRs}

The quantum routing property of single photons is characterized by
the transmission coefficient $T_{a(b)}=|t_{a(b)}|^2$ and reflection
coefficient $R_{a(b)}=|r_{a(b)}|^2$. It is easily seen from
Eq.~(\ref{eq:tr}) that these coefficients can be engineered.

\subsection{Photonic routing with a single WGR}

As a comparison, we first discuss the routing capability of a single
WGR, by assuming $V_{2a}=V_{2b}=0$. In Fig.~2, we plot $T_{a(b)}$
and $R_{a(b)}$ as a function of
$\Delta\omega(\Delta\omega=E_{k}-\omega_{c1})$ for different
backscattering strengths without any dissipation. It is seen that

(i) For $h_1=h=0$, i.e, single bosonic routing case, one can see
that $R_{b}=1$ and $R_{a}=T_{a}=T_{b}=0$ for $\Delta\omega=0$. This
indicates that, in the resonant case, the single photon is
transferred completely into the left of the waveguide-b along with
the sidewall of the resonator. In this case, the counterclockwise
mode of the WGR is not excited and thus the incident photon cannot
transport to the left of waveguide-a and the right of waveguide-b.

(ii) With the intermode interaction between the bosonic modes in the
GWR, i.e, $h_1=h\neq 0$, $R_{a}$ and $T_{b}$ are out of zero. In
this case, the photon is routed into four outports, and the
probabilities are allocated in four ports.

\begin{figure}[htbp]
\includegraphics[width=8cm,height=6cm]{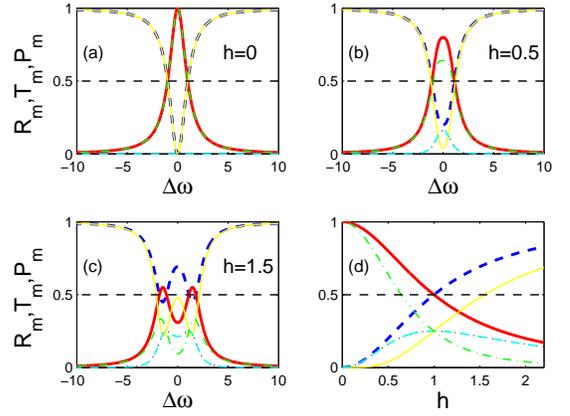}
\caption{(Color online) The transmission $T_{m}$ (solid yellow $T_a$
and dashed cyan $T_b$ ), reflection $R_{m}$ (dotted black $R_a$ and
dash dotted green $R_b$), and $P_m$ (thick dashed blue $P_a$ and
thick solid red $P_b$) for different backscattering strengths (a)
$h=0$, (b) $h=0.5$, (c) $h=1.5$. (d) as a function of backscattering
$h$ for $\Delta\omega=0.1$. Other parameters are
$\gamma_1=\gamma_2=0$, $\omega_{c}=2\pi\times6.0446
$MHz~\cite{Shen0}, $V_{1a}=V_{2a}=1$. For convenience, all the
parameters except the distance are in units of $\omega_{c}$.}
\end{figure}

To specifically investigate how the backscattering strength $h$
influence the routing between two channels, we plot $P_a$ and $P_b$
as a function $\Delta\omega$ for different backscattering strength
$h$ in Fig.~2, with $V_{1a}=V_{2a}=V_{a}=1$. Here, $P_a =T_a+R_a$
and $P_b =T_b+R_b$ represent the probabilities of finding the
incident single photons in the waveguide-a and the waveguide-b,
respectively. Interesting, a large region of $P_{b}>0.5$ in Fig.~2
(a) can emerge near the resonance point of the scattering spectra,
in contrast to the usual single-emitter routing wherein
$P_{b}\leq0.5$~\cite{Zhou,Lu,Lu2,Yan}. Physically, this may result
from the special construction of the WGR, wherein the cyclic modes
can couple and redirect the photons in waveguide-a into waveguide-b
in the opposite direction, along with the sidewall of the resonator
(see Fig.~1). Also, with the increase of the backscattering strength
$h$, $R_b$ decreases rapidly and others increase to cross a dot with
equal probabilities, and the region of $P_{b}>0.5$ decreases
gradually to zero around the point $h_1=h=1$, as shown in Fig.~2
(d). This may be due to the fact that, as increasing the
backscattering strength, the counterclockwise propagation is favored
and the transfer to waveguide-b is suppressed.

\subsection{Photonic routings with double WGRs}

Then we investigate the single-photon scattering properties in four
ports in the double-resonator case. For simplicity, we assume that
$\omega_{c1}=\omega_{c2}=\omega_{c}$, $h_{1}=h_{2}=h$,
$V_{1a}=V_{2a}=V_{a}$, and $V_{1b}=V_{2b}=V_{b}$. We plot $T_{a(b)}$
and $R_{a(b)}$ as a function of
$\Delta\omega(\Delta\omega=E_{k}-\omega_{c})$ for different
backscattering strengths under the distance $d=2$ of two WGRs in
Fig.~3. Similar to the single-resonator case, a large window of
$P_{b}>0.5$ appears near the resonance point, and the window
decreases with the increase of the backscattering strength.
Specifically, around the point $h=2$, the probabilities allocated in
the four ports meet at a equal-value dot, and $P_{b}$ reduces to
0.5.

\begin{figure}[htbp]
\includegraphics[width=8cm,height=6cm]{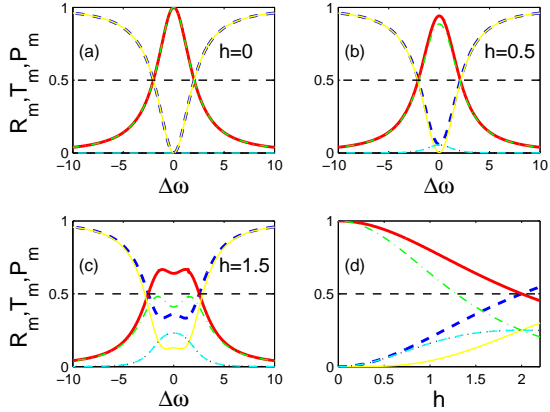}
\caption{(Color online)  The transmission $T_{m}$ (solid yellow
$T_a$ and dashed cyan $T_b$ ), reflection $R_{m}$ (dotted black
$R_a$ and dash dotted green $R_b$), and  $P_m$ ( thick dashed blue
$P_a$ and thick solid red $P_b$) for different backscattering
strengths: (a) $h=0$, (b) $h=0.5$, (c) $h=1.5$. (d) as a function of
backscattering $h$ for $\Delta\omega=0$. Other parameters are
$\gamma_1=\gamma_2=0$, $V_a=V_b=1$ and $d=2$ (in unit of
$1.6\times10^{-10}\upsilon_{g}/\omega_{c}$).}
\end{figure}

\begin{figure}[htbp]
\includegraphics[width=8cm,height=5cm]{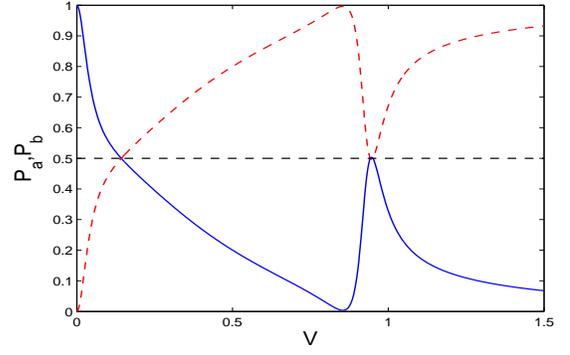}
\caption{(Color online) $P_a$ (solid blue) and $P_b$ (dashed red) as
a function of waveguide-WGR coupling strength. These parameters are
$\gamma_1=\gamma_2=0$, $h=0.5$, $\Delta\omega=0.42$, $d=1$ (in unit
of $1.6\times10^{-9}\upsilon_{g}/\omega_{c}$).}
\end{figure}

The coupling strengths between waveguides and WGRs also play a
crucial role for the quantum routing property. We plot $P_{a}$ and
$P_{b}$ as a function of the coupling strength $V=V_{a}=V_{b}$ under
the non-resonance regime (e.g. $\Delta\omega=0.42$) in Fig.~4. It is
shown that, when the coupling strengths are shut down the incident
single photon will transmit completely in the waveguide-a. When the
coupling strengths are turned on, the probabilities of the single
photon are allocated in two waveguides and a large region of high
transfer appears as the incident single photon transfer from the
waveguide-a to the waveguide-b.

\begin{figure}[htbp]
\includegraphics[width=8cm,height=6cm]{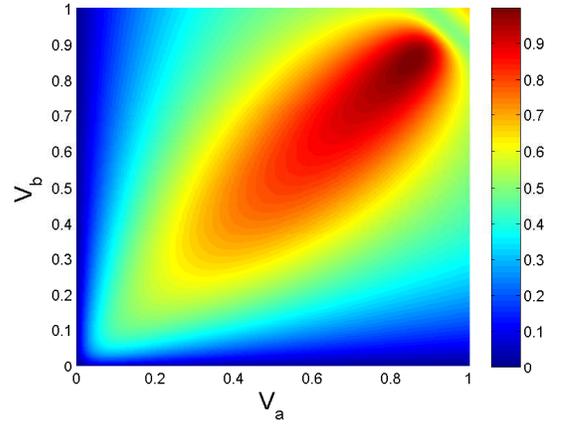}
\caption{(Color online) $P_b$ as functions of waveguide-WGR coupling
strengths $V_a$ and $V_b$. These parameters are the same as Fig.~4.}
\end{figure}

\begin{figure}[htbp]
\includegraphics[width=8cm,height=6cm]{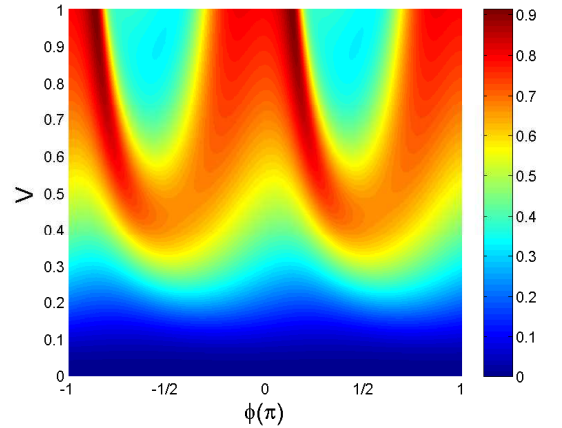}
\caption{(Color online) $P_b$ as functions of phase shift $\phi(k)$
and coupling strength $V$. These parameters are
$\gamma_1=\gamma_2=0$, $h=1$, $\Delta\omega=0.42$.}
\end{figure}

To gain a deeper insight into the dependence of the inter-channel
transport on the waveguide-WGR couplings, Fig.~5 displays the
transfer rate $P_{b}$ versus the different couplings of two
waveguides. It is seen that, when $V_{a}$ and $V_{b}$ are both small
the incident photon cannot be redirected from the waveguide-a into
the waveguide-b completely. Increasing the couplings, a large window
of high transfer rate emerges. More interestingly, the transfer rate
$P_{b}$ can reach to a maximum value $P_{b}=0.996\approx1$ at the
equal value location around $V_a=V_b=0.85$, which means that the
perfect transfer can almost be realized.

Fig.~6 shows specifically how the transfer rate $P_{b}$ against the
coupling strength and phase shift $\phi$ for the typical parameters
$h=1$ and $\Delta\omega=0.42$. As seen, with increasing the coupling
strength a wide region of of $P_b>0.5$ appears, and even high
transfer rates of $P_b>0.9$ are located in the region. It is also
found that the transfer rate of $P_{b}$ is a periodic function of
the phase shift, and two small regions with low transfer rate
($P_b<0.5$) emerge near the standing wave antinodes with
$\phi=\pm\pi/2$.

\subsection{Engineering the Fano-like resonances}

Fano resonances are the popular resonant scattering phenomena of the
electric and magnetic transports in many condensed matter systems.
Due to the interference between the photonic waves, the Fano-like
resonance may also appear. To investigate such a phenomenon, we
examine the routing property of the single photon from the input
quantum channel to another one. Fig.~7 displays $P_a$ and $P_b$ as a
function $\Delta\omega$ for the different distance $d$ of the two
WGRs, with $V_a =V_b=1$ and $h=0.5$. Obviously, for $d=0$ the system
reduces to the single-resonator case without any interference of the
photonic wave. As a consequence, the Fano-like resonance doesn't
emerge. With the increase of $d$, the jiggling behavior emerges due
to multiple interference of waves, and the feature of Fano-like
resonance becomes more evidently. Physically, the WGR-1 (WGR-2) is
served as a delocalized (localized) channel for the single photon
passing through it, and thus the interference between these two
channels results in the asymmetric line
shape~\cite{Chen,Chen1,Chen2}. Certainly, increasing the distance of
two WGRs enhances the interference between the two channels, and
thus leads to more distinct Fano-line shapes.

\begin{figure}[htbp]
\includegraphics[width=8cm,height=6cm]{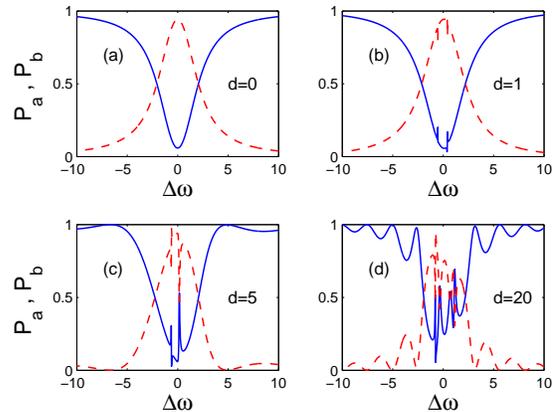}
\caption{(Color online) $P_a$ (solid blue) and $P_b$ (dashed red)
for different distance of the two WGRs (a) $d=0 $, (b) $d=1 $, (c)
$d=5 $ and (d) $d=20$ (in unit of
$1.6\times10^{-9}\upsilon_{g}/\omega_{c})$, respectively. Other
parameters are $\gamma_1=\gamma_2=0$, $V_a =V_b=1$ and $h=0.5$.}
\end{figure}

\section{Conclusions and Discussions}

Certainly, in the real experiments the dissipation of WGR is
inevitable during propagation, although it is significantly weaker
than the usual atom and single-mode resonator. Specifically, we plot
the transfer rate $P_{a}$, $P_{b}$ and loss $L$ ($L=1-P_{a}-P_{b}$)
for certain dissipations in Fig.~8, with $h=0.5$ and $d=1$. It is
seen that the altitude of $P_{b}$ decreases slightly with the
increase of dissipation. Therefore, the proposed double-WRG quantum
router should be robust for the further possible applications.

\begin{figure}[htbp]
\includegraphics[width=8cm,height=5cm]{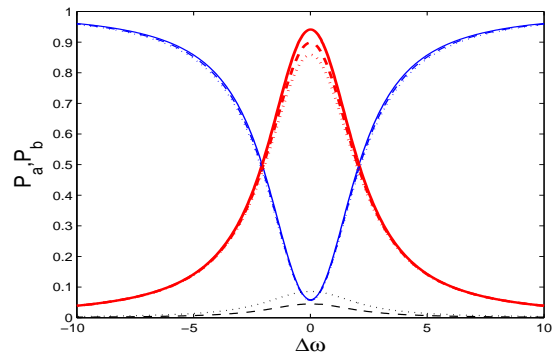}
\caption{(Color online)  $P_a$ (blue line), $P_b$ (red line) and
loss $L$ ($L=1-P_a-P_b$) (black line at the bottom of the plot)
versus $\Delta\omega$ for different dissipation in the WGRs
$\gamma_c=0$ (solid line), $\gamma_c=0.1$ (dashed line) and
$\gamma_c=0.2$ (dotted line). Other parameters are $h=0.5$, $V_a
=V_b=1$ and $d=1$ (in unit of
$1.6\times10^{-9}\upsilon_{g}/\omega_{c}$).}
\end{figure}

In summary, we have investigated the single-photon quantum routing
produced by a double-GRW quantum router. Using a full quantum
theory, the single-photon transmission and reflection amplitudes
were analytically obtained. By numerical method, we analyzed the
relevant transport properties in detail. It is found that, by
properly setting the relevant parameters, single-photon scattering
into four ports can be designed conveniently. Moreover, Fano-like
resonance, due to the quantum interference, is exhibited in the
scattering spectra. Our results showed that high routing capability
from input channel to another channel can be achieved near resonance
by only a single resonator, also implemented by adjusting the
distance of two WGRs and the coupling strengths between waveguide
and WGR, even in the nonresonance regime. Therefore, the proposed
system could be utilized as a robust quantum router.

\section*{Acknowledgments}

This work was supported by the National Natural Science Foundation
of China (Grant Nos. $11247032$ and U1330201) and the Natural Science Foundation
of Jiangxi (Grant No. $20151$BAB$202012$).

\end{document}